# Magnonic Charge Pumping via Spin-Orbit Coupling

**Authors:** Chiara Ciccarelli[1†], Kjetil M. D. Hals[2,3†], Andrew Irvine[1], Vit Novak[4], Yaroslav Tserkovnyak[5], Hidekazu Kurebayashi[1,6‡], Arne Brataas[2]* and Andrew Ferguson[1].

**Affiliations:**

[1]Cavendish Laboratory, University of Cambridge, Cambridge, CB3 0HE

[2] Department of Physics, Norwegian University of Science and Technology, NO-7491, Trondheim, Norway

[3]The Niels Bohr International Academy, Niels Bohr Institute, 2100 Copenhagen, Denmark

[4]Institute of Physics ASCR, v.v.i., Cukrovarnická 10, 162 53 Praha 6, Czech Republic

[5]Department of Physics and Astronomy, University of California, Los Angeles, California 90095, USA

[6]PRESTO, Japan Science and Technology Agency, Kawaguchi 332-0012, Japan

[†]These authors contributed equally.

[‡]Present address: London Centre for Nanotechnology, University College London, London WC1H 0AH, U.K. and Department of Electronic and Electrical Engineering, University College London, London WC1E 7JE, U.K.

*Correspondence to: arne.brataas@ntnu.no

**The interplay between spin, charge, and orbital degrees of freedom has led to the development of spintronic devices like spin-torque oscillators, spin-logic devices, and spin-transfer torque magnetic random-access memories. In this development spin pumping, the process where pure spin-currents are generated from magnetisation precession [1], has proved to be a powerful method for probing spin physics and magnetisation dynamics [1, 2, 3, 4, 5, 6, 7]. The effect originates from direct conversion of low-energy quantised spin-waves in the magnet, known as magnons, into a flow of spins from the precessing magnet to adjacent normal metal leads. The spin-pumping phenomenon represents a convenient way to electrically detect magnetisation dynamics [2, 3, 4, 5, 6, 7], however, precessing magnets have been limited so far to pump pure spin currents, which require a secondary spin-charge conversion element such as heavy metals with large spin Hall angle [5, 6, 7] or multi-layer layouts [8] to be detectable. Here, we report the experimental observation of charge pumping in which a precessing ferromagnet pumps a charge current, demonstrating direct conversion of magnons into high-frequency currents via the relativistic spin-orbit interaction. The generated electric current, differently from spin currents generated by spin-pumping, can be directly detected without the need of any additional spin to charge conversion mechanism and contains amplitude and phase information about the relativistic current-driven magnetisation dynamics. The charge-pumping phenomenon is generic and gives a deeper understanding of the recently observed spin-orbit torques, of which it is the reciprocal effect and which currently attract interest for their potential in manipulating magnetic information. Furthermore, charge pumping provides a novel link between magnetism and electricity and may find application in sourcing alternating electric currents.**

A flow of spin angular momentum without an accompanying charge current is called a pure spin current. A simple way to generate pure spin currents is via spin-pumping [1]. The phenomenon originates from direct conversion of low-energy quantised spin-waves in the magnet, known as magnons, into a flow of spins from the precessing magnet to adjacent normal metal leads. The reciprocal effect, in which a spin current is able to excite magnetisation dynamics, is known as the spin-transfer torque. In this case spin-angular momentum is transferred from the carriers to the magnet, applying a torque to the magnetisation[9]. This pair of reciprocal effects underlies much of the progress in spintroincs to date.

The spin-orbit coupling provides an efficient route to electrically generate magnetic torques from orbital motion, i.e., from an electric current (Fig. 1a,c) [10, 11, 12, 13, 14, 15, 16, 17]. These relativistic spin-orbit torques (SOTs) exist in ferromagnets with broken spatial inversion symmetry. They have been reported in (Ga,Mn)As, a material with a broken bulk inversion symmetry [18, 19, 20, 21] as well as heterostructures comprising ferromagnetic metals [22, 23, 24, 25, 26, 27]. The SOT has been observed to have both field-like [18, 22] and damping-like [21, 24, 25] contributions. Differently from non-relativistic spin-transfer torques, SOTs do not rely on a secondary element that spin-polarises the currents: rather a spin-polarisation results from the carrier velocity. Despite showing promise for magnetic memory applications, the understanding of SOTs is, however, still immature and a further development of the field requires improved theoretical models and experimental techniques to reveal their full complexity. The Onsager reciprocity relations [28] imply that, as for spin-pumping/spin transfer torque, there exists a reciprocal phenomenon of the SOT, namely, charge pumping generated from magnetisation precession (Fig. 1b,d) [14, 29].

The underlying physics of charge pumping is direct conversion of magnons into charge currents via the spin-orbit coupling. We will therefore refer to this process as *magnonic charge pumping*. Any external force that drives magnetisation precession can generate magnonic charge pumping. Examples of potential driving forces are magnetic fields, alternating currents, thermal gradients, or circularly polarised light pulses. Magnonic charge pumping can be a favourable alternative to spin pumping for detection of magnetisation dynamics, because the effect does not require an additional conversion mechanism to be measureable. Moreover, charge pumping contains information about the SOTs, and therefore opens the door for a novel experimental technique to explore these relativistic torques. Since the coefficients that describe the SOT are related to those that describe charge pumping, via the Onsager relations, it is possible to experimentally measure the amplitude and symmetry of the spin-orbit torque, in order to determine the expected charge-pumping signal. In our experiment, we do this and compare the result to the experimentally measured charge-pumping signal.

A simple explanation of magnonic charge pumping can be found from the Hamiltonian

$$H = \mathbf{p}^2/2m + \mathbf{p}\cdot\mathbf{\Lambda}\cdot\boldsymbol{\sigma} + \Delta\mathbf{m}\cdot\boldsymbol{\sigma},  \qquad (1)$$

where $\boldsymbol{\sigma} = (\sigma_1, \sigma_2, \sigma_3)$ is the carrier's spin operator represented by the Pauli matrices $\sigma_i$, $\mathbf{p}$ is the momentum operator, $\Delta$ is the exchange splitting and $\mathbf{m}$ is the unit vector in the direction of the magnetisation. The second term in the Hamiltonian (1) represents the spin-orbit coupling, where the matrix $\mathbf{\Lambda}$ parameterises the spin-orbit coupling. The velocity operator resulting from the Hamiltonian (1) is

$$\mathbf{v} = \partial H/\partial \mathbf{p} = \mathbf{p}/m + \mathbf{\Lambda}\cdot\boldsymbol{\sigma}.  \qquad (2)$$

The last term in Eq. (2) is the anomalous term, which mediates a coupling between spin and momentum. In ferromagnets, excitations of magnons result in a net non-equilibrium spin accumulation $\delta\langle\boldsymbol{\sigma}\rangle(t)$ due to the exchange interaction, yielding an average velocity response $\delta\langle\mathbf{v}\rangle(t) = \mathbf{\Lambda}\cdot\delta\langle\boldsymbol{\sigma}\rangle(t)$ which produces an alternating current density $\mathbf{j} \sim \mathbf{\Lambda}\cdot\delta\langle\boldsymbol{\sigma}\rangle(t)$. Since the magnon frequencies are low compared to the exchange splitting, the spin-density response is proportional to the rate of change $\partial\mathbf{m}/\partial t$ of the magnetisation, i.e., $\delta\langle\boldsymbol{\sigma}\rangle(t) \sim \partial\mathbf{m}/\partial t$. Consequently, the induced current density is also proportional to $\partial\mathbf{m}/\partial t$, where the coefficient of proportionality is directly related to the spin-orbit coupling matrix: $\mathbf{j} \sim \mathbf{\Lambda}\cdot\partial\mathbf{m}/\partial t$.

We chose compressively strained (Ga,Mn)As on GaAs as the material in which to demonstrate magnonic charge pumping. (Ga,Mn)As is indeed characterized by crystal inversion asymmetry, which together with strain leads to easily identifiable SOTs with both Rashba and Dresselhaus symmetry [30]. Furthermore, the use of (Ga,Mn)As avoids the complexity associated with a competing torque originating in the spin Hall effect, that is present in the layered metal systems [20, 31]. The symmetry of strained (Ga,Mn)As is described by the crystallographic point group $C_{2v}$, where the two-fold symmetry axis is perpendicular to the epilayer [30]. In the frame of reference where x' is along the crystallographic direction [110], z' is along the two-fold symmetry axis and y' is perpendicular to x' and z', $\mathbf{\Lambda} \sim i\sigma_2$ and $\Lambda$~$\sigma_1$ parameterise the Rashba and Dresselhaus spin-orbit coupling, respectively, and the induced alternating current density is

$$\mathbf{j}^{(r)} = \Lambda_D^{(r)}\,\sigma_1\cdot\partial\mathbf{m}_\parallel/\partial t - i\Lambda_R^{(r)}\,\sigma_2\cdot\partial\mathbf{m}_\parallel/\partial t. \qquad (3)$$

Here, $\mathbf{m}_\parallel = (m_{x'}, m_{y'})$ denotes the in-plane component of the magnetisation, and the parameters $\Lambda_R^{(r)}$ and $\Lambda_D^{(r)}$ characterise the strength of the charge current pumped magnonically via the Rashba and Dresselhaus spin-orbit coupling, respectively. The current density in Eq. (3) is reciprocal to the field-like SOT $\boldsymbol{\tau}$~$\mathbf{m}\times\mathbf{h}^{so}$, where $\mathbf{h}^{so}$ is the effective SOT field induced by an applied current density $\mathbf{J}$. The SOT field consists of terms with Rashba and Dresselhaus symmetry, i.e., $\mathbf{h}^{so} = h_D\sigma_1\cdot\mathbf{J}_\parallel + ih_R\sigma_2\cdot\mathbf{J}_\parallel$, where the parameters $h_R$ and $h_D$ are linked via the reciprocity relations to $\Lambda_R^{(r)}$ and $\Lambda_D^{(r)}$, respectively.

The terms in Eq. (3) represent reactive charge-pumping processes because they are even under time reversal. In addition, there are dissipative contributions to the magnonic charge pumping, which are related via the reciprocity relations to the anti-damping SOT. The in-plane component of the dissipative current is (see Supplementary Information for a detailed derivation)

$$\mathbf{j}^{(d)} = \Lambda_R^{(d)}\,\mathbf{m}_\parallel\,\partial m_{z'}/\partial t + \Lambda_D^{(d)}\,\sigma_3\cdot\mathbf{m}_\parallel\,\partial m_{z'}/\partial t, \qquad (4)$$

where the phenomenological parameters $\Lambda_R^{(d)}$ and $\Lambda_D^{(d)}$ characterise dissipative charge pumping by Rashba and Dresselhaus SOC, respectively.

When the magnetisation precesses with frequency $\omega_0$ and amplitude A, there is a reactive contribution to the pumped current oscillating at the same frequency with an amplitude of $j_\omega^{(r)} = A\omega_0\Lambda_{R,D}^{(r)}$. The polar plot in Fig. 1d shows the symmetry of the Rashba and Dresselhaus contributions to the pumped current for different directions of the magnetisation. Fig. 1c illustrates the symmetry of the reciprocal effect and shows the direction of the reactive components of the Rashba and Dresselhaus SOT fields for different directions of the applied current. There is also a direct current induced by the magnetisation precession (see Supplementary Information). However, its value is small since it is second order in the precession amplitude and proportional to the Gilbert damping constant $\alpha_G$ and will not be discussed further.

Fig. 2a shows a schematic of the measuring apparatus. In order to drive magnetisation precession via the SOT, a microwave current is passed through a micro-bar patterned from an epilayer with a nominal 9 % Mn concentration. During magnetisation precession, frequency mixing between the alternating current and the oscillating magneto-resistance leads to a time-independent voltage, $V_{dc}$ [20]. Using $V_{dc}$, we experimentally determine the components of the SOT, introducing a rotated reference frame where x is along the bar (current) direction and z is perpendicular to the epilayer (Fig. 2). The angle θ refers to the mean position of the magnetisation in the x-y plane and is measured from the x-axis. We focus our experiments on the two bar directions [100] and [010], because the SOT field components $h_{so}^x$ and $h_{so}^y$ then originate purely from the field-like SOTs which have symmetries that respectively resemble the Dresselhaus and Rashba spin-orbit interactions (Fig. 1c). Fig. 2b shows the derivative of the rectified voltage $(dV_{dc}/dB)B_{mod}$ for a bar oriented along the [100] direction when an in-plane magnetic field B is swept through the ferromagnetic resonance condition. The position of the resonance as a function of the field direction follows the modified Kittel's formula for an in-plane magnetised material with an additional uniaxial anisotropy [32]. The SOT-field $\mathbf{h}_{so}$ can be directly extracted from the angle dependence of the anti-symmetric and symmetric parts of the resonance, and the coefficients are summarised in the following table. The SOT field components $h_{so}^x$ and $h_{so}^y$ correspond to the coefficients $h_D$ and $h_R$ introduced earlier, while the angle-dependent $h_{so}^z$ terms represent the anti-damping contribution. In accordance with a trend that we previously observed, the presently used material has a weaker SOT than in the case of lower Mn concentration [20].

**Table 1.** The coefficients of SOT measured for samples with current along the [100] and [010] directions, normalised to a current density of $10^6$ Acm$^{-2}$. All values are in (μT). The first order (sinθ and cosθ) harmonic components of $h_{so}^z$ are extracted from fits to the experimental data.

|  | $\mu_0 h_{so}^x$ | $\mu_0 h_{so}^y$ | $\mu_0 h_{so}^z$ sinθ term | $\mu_0 h_{so}^z$ cosθ term |
|---|---|---|---|---|
| [100] | -6.1 | -8.7 | 8.5 | -13.6 |
| [010] | 5.2 | -5.5 | -5.5 | -6.9 |

In addition, the magnonic charge pumping induces an alternating voltage, $V_\omega$, across the bar, which we measure with a field modulation lock-in technique (see Supplementary Information). Fig. 2c shows the derivative of the amplitude of the microwave voltage across the sample, $(dV_\omega/dB)B_{mod}$, as the magnetic field is swept along different in-plane directions. At ferromagnetic resonance, a resonance also appears in $V_\omega$, which indicates that a microwave electrical signal is generated within the sample by the precessing magnetisation.

Magnonic charge pumping is proportional to the rate of change of the magnetisation, hence the induced microwave amplitude should be linearly dependent on the precessional amplitude. To check this characteristic, we measure the voltage $V_\omega$ as a function of the precessional amplitude A for a fixed direction of the magnetic field. The amplitude is controlled by the value of the applied microwave current. Fig. 2d clearly demonstrates a linear dependence on the amplitude. This excludes the possibility that $V_\omega$ originates from the mixing between the microwave current and the modulated resistance during precession, because such higher-order terms depend non-linearly on the amplitude (see Supplementary Information).

Next, we demonstrate that the measured signal is reciprocal to the SOT. To this end, we model the charge pumping by Eqs. (3)-(4) (see Supplementary Information for further details). Using the Onsager reciprocity relations, the measured SOT fields $h_{so}^y$ and $h_{so}^x$ determine the values of $\Lambda_R^{(r)}$ and $\Lambda_D^{(r)}$, respectively, while the measured $h_{so}^z$ component determines $\Lambda_R^{(d)}$ and $\Lambda_D^{(d)}$. The expression for $\partial \mathbf{m}/\partial t$ is found from the solution of the LLG equation. The resulting voltage signal across the bar is given by the total current pumped along the bar direction multiplied by the resistance. In Fig. 3a-b, we plot the magnitude of the symmetric and anti-symmetric components of the integrated resonances with respect to the field direction. The theoretical curves are represented by the continuous lines and show agreement with the experimental data in both symmetry and amplitude. This verifies that the measured voltage signal satisfies its reciprocal relationship to the SOT. The different symmetries found for the [100] and the [010] bar directions further confirm the crystal, thus SOC-related, origin of the effect and exclude the Oersted field and artefacts in the measuring set-up as possible origins. Also, a variation of the impedance matching following the ac change in magnetic susceptibility during precession cannot justify the resonance in $V_\omega$ as in this case the symmetry would be dominated by the symmetry of the anisotropic magneto-resistance (refer to the Supplementary Information). The slight discrepancy between the experimental and theoretical curves arises from higher-order harmonics in the phenomenological expansion of the pumped current. Such higher-order features have also been observed in the SOT [25]. To allow comparison of the magnonic charge pumping between different materials, we renormalise the pumped current density by the saturation magnetisation, frequency, and precessional amplitude. For our (Ga,Mn)As samples, we find a magnitude of 600 µA·cm$^{-2}$/T·GHz for the [100] direction and 240 µA·cm$^{-2}$/T·GHz for the [010] direction. In (*20*) the authors reported fluctuations of 30% in the magnitude of the SOT for samples of the same material. Likewise, the magnitude of the charge pumping is expected to be sample-dependent, although its symmetry is only determined by the crystalline orientation of the bar, as also shown in the Supplementary Information.

In conclusion, we have demonstrated direct conversion of magnons into high-frequency currents via the spin-orbit coupling. While we chose the ferromagnetic semiconductor (Ga,Mn)As, magnonic charge pumping is also predicted in layered systems like Pt/Co/Al$_2$O$_3$ and can be quantitatively analysed within the same Onsager framework we provide. In these metallic systems, we expect a large, room-temperature charge pumping effect, the investigation of which will help distinguish between spin Hall and spin-orbit torques.

**Methods and Materials:** <u>Materials</u>: The 18 nm thick (Ga$_{0.91}$,Mn$_{0.09}$)As epilayer was grown on a GaAs [001] substrate by molecular beam epitaxy. It was subsequently annealed for 8 hours at 200 ºC. It has a Curie temperature of 179 K; a room temperature conductivity of 414 $\Omega^{-1}$cm$^{-1}$, which increases to 544 $\Omega^{-1}$cm$^{-1}$ at 4 K; and a saturation magnetisation of 70.8 emu·cm$^{-3}$.

<u>Devices</u>: Two terminal microbars are patterned in different crystal directions by electron beam lithography and have dimensions of 4 µm × 40 µm.

<u>Experimental procedure</u>: A 7 GHz microwave signal with a source power of 18 dBm is transmitted to an impedance matching circuit made by a 4-finger interdigitated capacitor and a λ/2 micro-strip resonator patterned on a low loss printed circuit board and reaches the (Ga,Mn)As bar, which is wire-bonded between the resonator and the ground plane. SOT excites magnetic precession as an external field is swept in the plane of the device. The microwave voltage generated in the (Ga,Mn)As bar by magnonic charge pumping is transmitted via a directional coupler to an amplifier and mixer, from which we measure the amplitude of the voltage. Low frequency (222 Hz) field modulation with an amplitude of 3.3 mT is adopted, along with lock-in detection, to remove the charge pumping signal from the reflected microwave signal. When driven at its fundamental frequency (7 GHz), there is a node of electric field at the centre point of the resonator and it is possible to incorporate a bias-tee by simple wire-bonding. This allows measuring the rectification voltage across the bar. All the measurements are performed at a temperature of 30 K.

**Acknowledgments**

AJF acknowledges support from a Hitachi research fellowship and CC from a Junior research fellowship at Gonville and Caius College. VN acknowledges MSMT grant Nr. LM2011026.

**Author contributions**

KH and AB developed the theory and suggested the experiment. CC and AJF developed the experimental technique and performed the experimental work. VN grew the materials. ACI performed the nanofabrication. CC, KH, AB & AJF wrote the manuscript. All authors discussed the results and commented on the paper.

**Competing Financial Interests**

The authors declare no competing financial interests.

**Fig. 1**. (a) A charge current through (Ga,Mn)As results in a non-equilibrium spin polarisation of the carriers, which exchange-couples to the magnetisation and exerts a torque. The effect is induced by the spin-orbit coupling, which mediates the transfer of orbital momentum to spin angular momentum. An alternating current generates a time varying torque, which drives magnetic precession resonantly when a magnetic field is applied. (b) The reciprocal effect of (a). Magnetisation precession leads to a non-equilibrium spin concentration, which is converted into an alternating charge current by the spin-orbit coupling. (c) Polar plot illustrating the direction of the effective magnetic field induced by a charge current along different crystal directions. The Rashba ($h_R$) and Dresselhaus ($h_D$) spin-orbit coupling contributions are indicated by the red and blue arrows, respectively. (d) Polar plot illustrating the direction of the charge current pumped by magnetisation precession around different

crystal directions. The Rashba ($j_R$) and Dresselhaus ($j_D$) contributions are indicated by the red and blue arrows, respectively.

**Fig. 2.** (a) Schematics of the measuring set-up. A 7 GHz microwave signal (red arrow) is launched towards a (Ga,Mn)As bar via an impedance matching circuit. The microwave current passed through the bar excites magnetisation precession via SOT when an in-plane magnetic field B is swept through the resonance. The orientation of the field is defined with respect to the bar direction, as shown on the Cartesian plot. The microwave voltage generated in (Ga,Mn)As by magnonic charge pumping (blue arrow) is transmitted through the same impedance matcher to the microwave circuitry, where the amplitude of the signal is amplified and detected. A low-frequency lock-in field-modulation technique is used: a 3.3 mT oscillating magnetic field $B_{mod}$ is applied at 45° from the bar direction. A directional coupler separates the incoming signal used to excite magnetic precession (red arrow) from the outgoing signal generated both by magnonic charge pumping and the microwave signal reflected from the circuit (blue arrow). The impedance matching circuit also includes a bias tee that allows the rectified voltage along the bar to be measured. (b) Derivative of the rectified voltage along the a [100] oriented bar, $(dV_{dc}/dB)B_{mod}$, measured by field modulation lock-in technique as the magnetic field is swept along different in-plane directions. (c) Derivative of the microwave voltage along a [100] oriented bar, $(dV_{\omega}/dB)B_{mod}$, induced by magnonic charge pumping for the same field directions as in (b). (d) Amplitude of the microwave voltage $V_{\omega}$ as a function of the precessional amplitude A. The value of A in mrad is obtained from the amplitude of the rectified voltage $|V_{dc}|=|I|\Delta RA/2$, where I is the microwave current passed through the bar and $\Delta R$ is the anisotropic magneto-resistance coefficient.

**Fig. 3.** Symmetric (red circles) and anti-symmetric (black squares) components of the integrated resonances shown in Fig. 2c for different directions of the external field B. (a) shows the results obtained for a bar oriented along the [100] crystal direction, while (b) for a bar along the [010] direction. The data is compared to the theoretical curves (symmetric and anti-symmetric contributions are equal and are shown by a unique blue line).

FIGURE 1

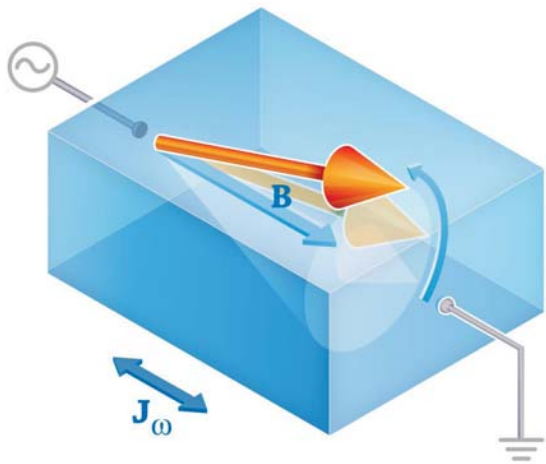

a.

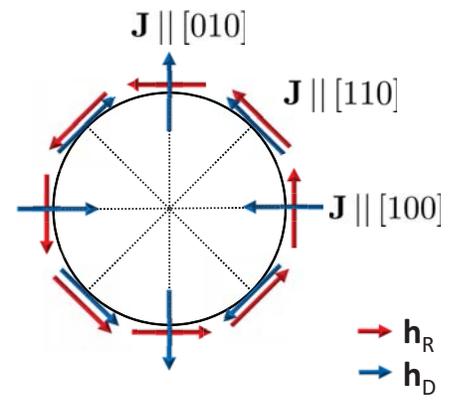

c.

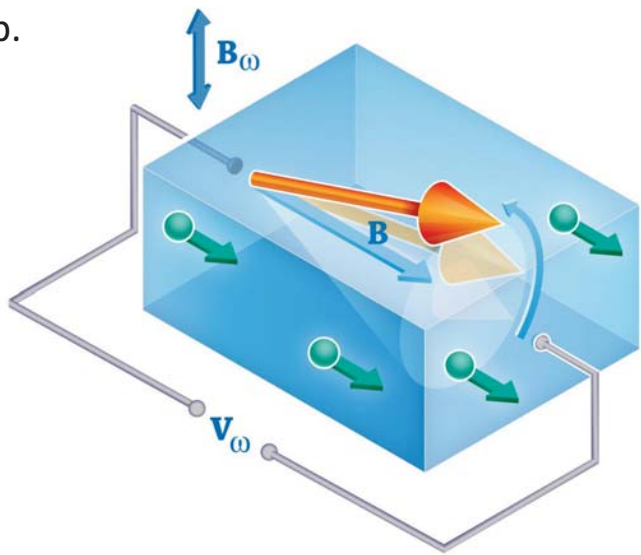

b.

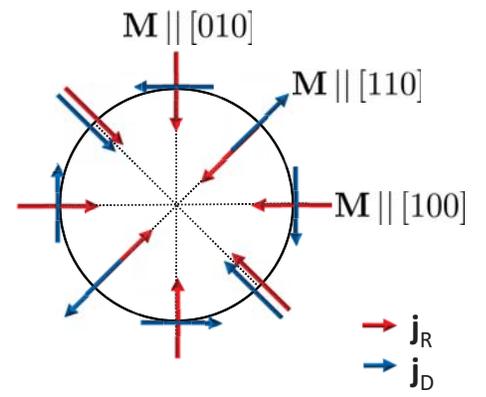

d.

FIGURE 2

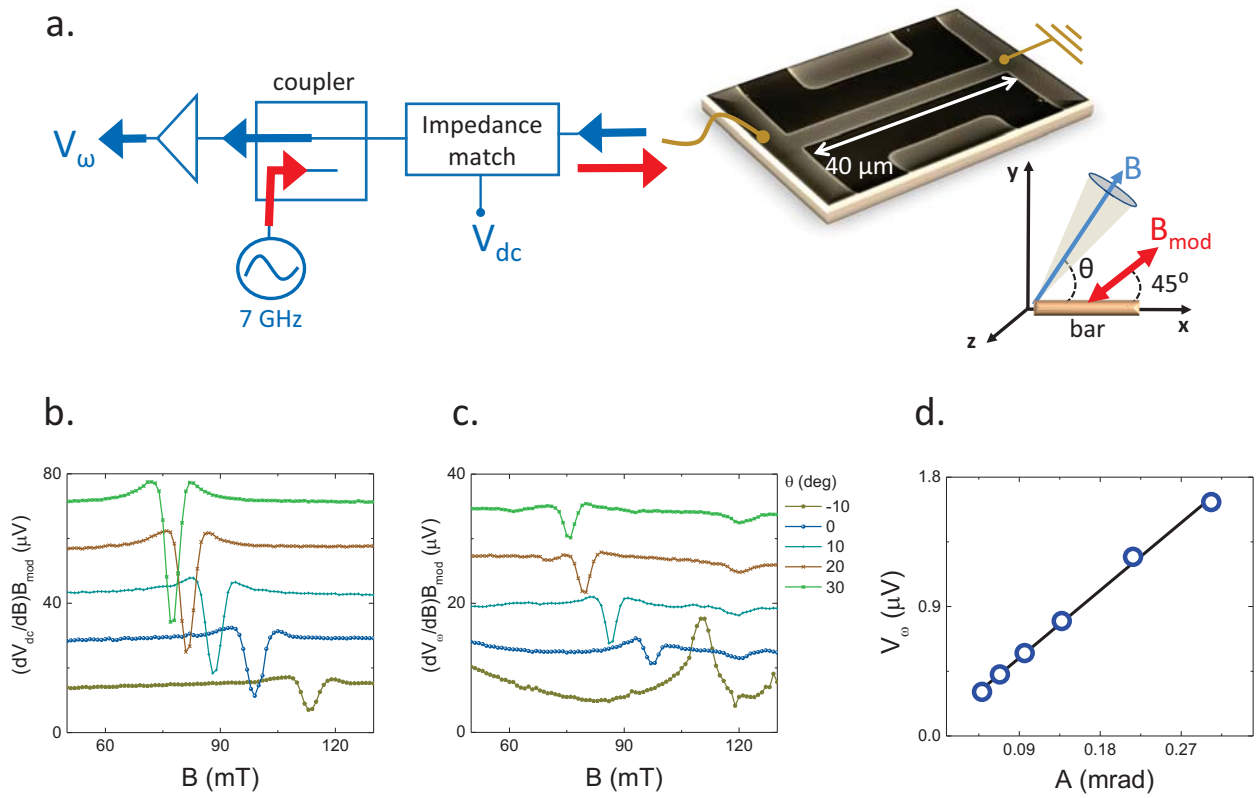

FIGURE 3

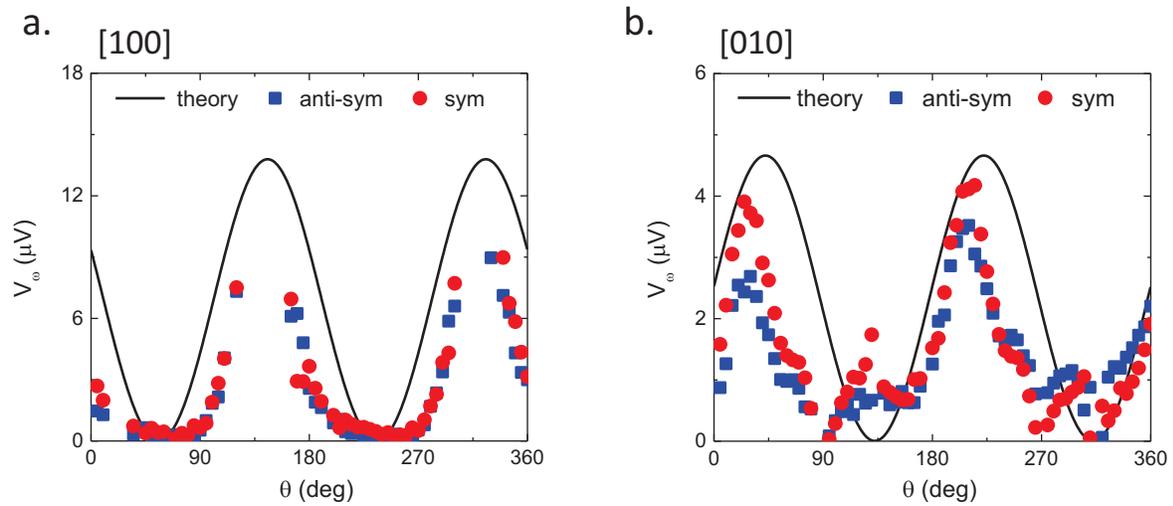